\documentclass[final,5p,times,twocolumn]{elsarticle}
\usepackage{amsmath, amssymb}
\usepackage{graphicx}
\usepackage{hyperref}
\usepackage{siunitx}
\usepackage{xcolor}
\usepackage{stfloats} 
\journal{Current Opinion in Colloid and Interface Science}

\begin{document}

\begin{frontmatter}

\title{Advanced Modelling Methodologies for Anisotropic Magnetic Colloids}

\author{Jorge L. C. Domingos}
\address{MMML Lab, Department of Physics, University of Latvia, Jelgavas iela 3, Riga LV-1004, Latvia}

\begin{abstract}
Anisotropic magnetic colloids with permanent dipole moments exhibit rich field-responsive behavior arising from the interplay between particle geometry, dipolar interactions, and external driving. Modeling these systems remains challenging due to the long-range nature of dipolar forces, geometric anisotropy, dipole--particle misalignment, and the complexity of implementing anisotropic steric interactions. This review discusses particle-based numerical strategies to model such systems, including single-site, multi-bead, shifted-dipole, and multicore representations. We analyze how different levels of description capture key physical mechanisms, from steric constraints and directional binding to internal magnetic structure and nonequilibrium dynamics. Particular emphasis is placed on dipole--particle misalignment as a control parameter that strongly affects interaction landscapes and self-assembly pathways. We also highlight recent machine learning approaches as emerging tools to construct effective interaction potentials and accelerate simulations. By comparing the main methodologies and their limitations, this review outlines current challenges and perspectives toward more predictive and efficient modeling of anisotropic magnetic colloids.
\end{abstract}

\end{frontmatter}

\section{Introduction}
Magnetic colloids with anisotropic shapes have attracted increasing attention due to their ability to form a wide range of field-responsive structures and dynamically tunable assemblies \cite{wang2024stimuli}. In particular, systems composed of particles carrying permanent magnetic dipole moments provide a well-defined framework to study how anisotropic dipolar interactions can be efficiently represented and simulated in particle-based models \cite{Singh2025SelfassemblyOD,836q-7vng,WOLFSCHWENGER2024115624}. Unlike magnetizable systems, where the dipole moment is induced by the applied field, permanently magnetized particles introduce an intrinsic directional interaction that remains present even in the absence of external driving, leading to qualitatively different aggregation pathways and steady states \cite{10.1063/5.0215545, PrezMarcos2025TimedependentCA}. In anisotropic particles, this complexity is further enriched when the dipole moment is not aligned with the main particle axis, giving rise to dipole-particle misalignment as an additional control parameter that can strongly affect both structure formation and dynamical response.

From a modeling perspective, these systems pose specific challenges. The long-range and anisotropic nature of dipolar interactions introduces significant numerical difficulties \cite{74l8-yfn7}, particularly when combined with geometric anisotropy, leading to complex energy landscapes \cite{doi:10.1021/acs.langmuir.4c03101,10.1063/5.0310699} that are highly sensitive to particle shape, dipole orientation, relative alignment, and collective effects. As a result, different levels of description have been developed, ranging from single-site anisotropic potentials \cite{Ellipsoidmodel2025,PhysRevE.110.054702} to multi-bead representations \cite{Domingos2025Soft,kantorivich2013} and coarse-grained models with distributed dipole moments. Each approach involves a trade-off between physical realism and computational cost, and their applicability depends strongly on the phenomena of interest. More recently, machine learning methods have emerged as a promising route to accelerate this modeling effort by enabling the construction of effective interaction potentials and data-driven coarse-grained representations for anisotropic systems \cite{Bupathy2025FromKT, Jin2024ImprovingML}.

This review focuses on numerical strategies used to model anisotropic magnetic particles with permanent dipoles. Emphasis is placed on how particle shape and dipole-particle misalignment are incorporated into particle-based simulations and on the corresponding modeling strategies used to represent these effects. We also discuss recent developments in data-driven approaches, particularly machine learning methods, as emerging tools to overcome current limitations in the simulation of anisotropic dipolar systems. The goal is not to provide an exhaustive overview of experimental realizations, but rather to highlight the methodological frameworks available to describe these systems and to identify their main advantages and limitations.
\section{Modelling Framework}

Magnetic colloids with permanent dipoles are typically modeled through a combination of dipolar and steric interactions. The dipole-dipole potential
\begin{equation}
U_{dd}(\mathbf{r}_{ij}) = \frac{\mu_0}{4\pi r_{ij}^3}
\left[
\boldsymbol{\mu}_i \cdot \boldsymbol{\mu}_j
- 3(\boldsymbol{\mu}_i \cdot \hat{\mathbf{r}}_{ij})
(\boldsymbol{\mu}_j \cdot \hat{\mathbf{r}}_{ij})
\right],
\end{equation}
is long-range, anisotropic, and decays as $r^{-3}$, favoring head-to-tail alignment while competing with thermal fluctuations and steric constraints.

Excluded volume effects are introduced via hard-core or soft repulsive potentials, such as dipolar hard-sphere (DHS), dipolar soft-sphere (DSS) \cite{domingos2017self}, or Stockmayer-type models \cite{BILOUS2026129210}. System behavior is primarily governed by a few dimensionless parameters. The dipolar coupling strength,
\begin{equation}
\lambda = \frac{\mu_0 \mu^2}{4\pi k_B T \sigma^3},
\end{equation}
measures the relative importance of dipolar interactions over thermal energy, controlling the transition from weakly interacting fluids ($\lambda \ll 1$) to strongly aggregated states ($\lambda \gg 1$). The volume fraction $\phi$ determines the role of many-body interactions, while external fields introduce an additional control parameter $\xi = \mu B / k_B T$, enabling field-driven assembly and dynamical regimes such as synchronization under time-dependent fields \cite{CAMACHO2025101903}.

From a computational perspective, the long-range nature of dipolar interactions poses a major challenge. For interactions decaying as $r^{-n}$ with $n \leq d$,  where $d$ is the spatial dimensionality of the system, convergence becomes non-trivial; in dipolar systems ($n=3$), three-dimensional systems lie at the borderline of long-range behavior \cite{allen1989computer}. As a result, naive truncation leads to artifacts, and accurate simulations typically rely on Ewald summation \cite{TOUKMAJI199673} or related techniques. Alternative methods, such as particle-particle particle-mesh (P3M) \cite{deserno1998mesh} and fast multipole methods \cite{darve2000fast}, offer improved scaling at the cost of additional complexity.

Within this framework, different simulation strategies are employed depending on the physical regime. Monte Carlo methods are commonly used for equilibrium sampling, while molecular dynamics (MD) provides access to time-resolved trajectories. In colloidal systems, overdamped Brownian or Langevin dynamics are often more appropriate, as they explicitly account for thermal fluctuations and viscous dissipation, and are essential to describe time-dependent processes such as aggregation and field-driven dynamics.

More recently, machine learning approaches have emerged as complementary tools to construct effective interaction potentials and accelerate force evaluations, enabling simulations at larger scales or longer times. Rather than replacing traditional methods, ML extends the modelling framework by alleviating current computational limitations.

Together, these elements define the minimal framework for modeling dipolar colloids. Extensions to anisotropic particles introduce additional complexity through particle shape and dipole orientation, as discussed in the following sections.
\section{Shape Anisotropy - The geometry factor}

Particle shape plays a central role in determining the collective behavior of magnetic colloids. 
Even when particles carry identical magnetic moments, elongation can qualitatively modify the effective dipolar interaction landscape. For sufficiently anisotropic particles, antiparallel side-by-side configurations may become energetically more favorable than the head-to-tail alignment typical of dipolar spheres, leading to the emergence of distinct equilibrium structures such as rings, carpets, or antiparallel aggregates depending on particle geometry and dipole orientation \cite{kantorivich2013}. 
From a modeling perspective, this implies that particle shape cannot be treated as a secondary correction to dipolar interactions, but must be explicitly incorporated into the interaction model in a consistent way. It is important to distinguish between different ways in which particle shape enters particle-based models of dipolar systems. In the simplest case, shape affects the interaction landscape indirectly through steric constraints, while the magnetic interaction is still described at the point-dipole level. In more detailed representations, shape is coupled to the spatial distribution of dipoles within the particle, for instance through multi-site or off-center dipole models. In this section, we focus on such particle-based descriptions where anisotropy arises from geometry rather than from material-dependent magnetization effects. In this context, geometric anisotropy is introduced through simplified particle representations.

\begin{figure*}[t]
    \centering
    \includegraphics[width=0.7\linewidth]{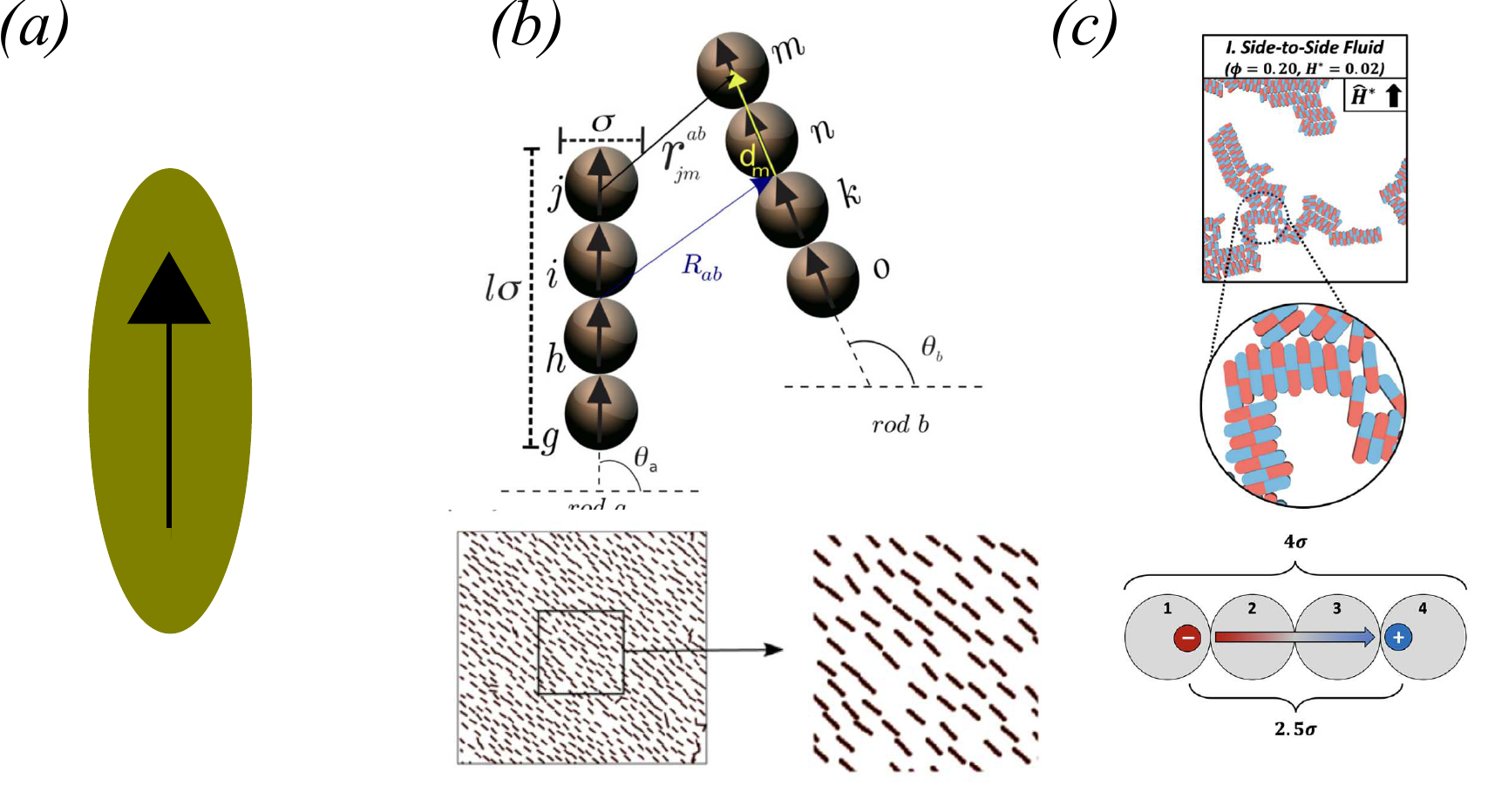}
    \caption{Representative modeling strategies for rod-like magnetic colloids. 
    (a) Gay-Berne single-site model, where anisotropy is incorporated through an effective orientation-dependent interaction potential. 
    (b) Multi-bead rod representation with distributed interaction sites, enabling an explicit description of particle geometry and dipolar interactions. Adapted from Ref. \cite{Domingos2025Soft} with permission from the Royal Society of Chemistry. 
    (c) Alternative multi-bead representation of an elongated anisotropic particle, illustrating a discrete-site approach to resolve shape and interactions. Adapted from Ref.\cite{Dorsey2025MagneticRods} with permission from the American Chemical Society.}
    \label{fig:shape1}
\end{figure*}

Single-site anisotropic potentials, such as ellipsoids (Fig.\ref{fig:shape1}(a)) or spherocylinders, provide a compact way to incorporate geometric anisotropy into theoretical and computational models. Steric interactions of ellipsoidal particles are commonly described using anisotropic potentials such as the Gay–Berne model \cite{gay1981modification}, whereas hard spherocylinder models provide a simple geometric description of elongated particles interacting via excluded volume \cite{Bolhuis1997HardRods}. 
These approaches reduce the particle description to a small set of orientational degrees of freedom, making them computationally efficient and well-suited for large-scale simulations. However, this reduction also limits their ability to capture fine geometric details, such as local curvature variations or non-uniform surface properties \cite{CASARELLA2024106221}, which can become relevant in systems where contact geometry strongly influences the interaction.

Beyond purely steric descriptions, recent studies have explored how particle shape can directly modify the magnetic interaction between anisotropic particles. In these approaches, the dipolar coupling between ellipsoids is derived from the magnetic disturbance field generated by uniformly magnetized particles and expressed through geometry-dependent interaction tensors \cite{suspellpsoid, Ellipsoidmodel2025}. 
For ellipsoidal particles, the field generated is obtained through a tensor that explicitly depends on particle aspect ratio and orientation, allowing the interaction to reflect the underlying geometry of the particle. Although these formulations are often applied to magnetizable particles with field-induced dipoles, the same framework can also describe permanently magnetized ellipsoids by prescribing the particle magnetization \cite{collecttiveellipsoids2025}. 
While more physically grounded, these tensor-based approaches are typically more complex to implement and computationally more demanding than single-site models, which may limit their use in large-scale simulations.

At the other end of the modeling spectrum, an alternative strategy consists of representing anisotropic particles as assemblies of interaction sites. 
In these coarse-grained multi-bead models, an anisotropic magnetic particle is represented as a set of spherical beads (Fig. \ref{fig:shape1}(b) ), allowing both excluded volume and magnetic interactions to be resolved at the level of individual interaction sites \cite{Domingos2025Soft,kantorivich2013, jorgeplos, jorgeprsoft}. 
This representation provides greater geometric flexibility, e.g. magnetic filaments \cite{D4NR01034E,D4SM00797B}, and facilitates the implementation of complex dipole distributions , including off-center dipoles, particle flexibility, and distributed magnetic moments. In coarse-grained multi-bead models, e.g., a magnetic rod of aspect ratio $l$ is represented as a linear chain of $l$ identical beads of diameter $\sigma$, each carrying a permanent dipole moment $\boldsymbol{\mu}_i$ aligned with the particle axis. The total dipole moment,
$\boldsymbol{\mu}_{\mathrm{rod}}=\sum_{i=1}^{l}\boldsymbol{\mu}_i$,
can be preserved across aspect ratios by scaling the bead moments as $\mu_i \sim 1/l$. 

This approach captures both intra-particle magnetic alignment and interparticle frustration, enabling configurations such as head–to–tail chains and side–by–side assemblies. While some of these configurations can already emerge in point-dipole models for sufficiently elongated particles, multi-bead representations enable a more realistic description of contact geometry and internal dipole distributions, allowing for a broader range of frustrated and anisotropic assemblies. However, this increased level of detail comes at a significant computational cost, as the number of interaction sites scales with the particle aspect ratio, leading to a substantial increase in the number of pairwise interactions that must be evaluated. As a result, multi-bead models are typically limited to moderate system sizes or require optimized algorithms to remain computationally feasible. The choice between single-site and multi-bead representations is therefore largely dictated by the level of geometric detail required to capture the relevant physical phenomena.

Related strategies include representing elongated particles as bonded hard disks with embedded opposite charges and simulating their dynamics using discontinuous molecular dynamics (Fig. \ref{fig:shape1}(c)), where stochastic field-induced impulses generate effective dipolar alignment without explicit torque integration \cite{Dorsey2025MagneticRods}. 
A similar methodology has been applied to square-shaped particles constructed from four bonded beads forming a rigid hard-colloid geometry \cite{Dorseypre2025}. These approaches provide an alternative route to incorporate anisotropy while avoiding the continuous integration of rotational degrees of freedom, offering computational advantages in specific regimes. However, they typically rely on simplified interaction rules, which may limit their applicability when detailed dynamical information is required.

A more detailed description of the dynamics for rod-like particles is typically achieved using Langevin dynamics with anisotropic friction coefficients \cite{Domingos2025Soft}. 
The translational and rotational motion of each rod is governed by
\begin{align}
M\frac{d\mathbf{v}}{dt}&=\mathbf{F}-\boldsymbol{\Gamma}_T\cdot\mathbf{v}
+\boldsymbol{\eta}_T(t),\\
I\frac{d\boldsymbol{\omega}}{dt}&=\mathbf{N}-\Gamma_R\boldsymbol{\omega}
+\boldsymbol{\eta}_R(t),
\end{align}
where $\mathbf{F}$ and $\mathbf{N}$ denote the total force and torque acting on the rigid body. 
The stochastic forces satisfy the fluctuation–dissipation relation
\[
\langle \boldsymbol{\eta}_\alpha(t)\boldsymbol{\eta}_\alpha(t') \rangle
=
2k_BT\,\boldsymbol{\Gamma}_\alpha\,\delta(t-t'), 
\qquad
\langle \boldsymbol{\eta}_\alpha(t)\rangle = 0,
\]
with $\alpha = T,R$. 
The contribution of each bead $m$ displaced $\mathbf{d}_m$ from the center of mass of the rod to the total magnetic torque is given by
\begin{equation}
\boldsymbol{\tau}_{m}= \boldsymbol{\mu}_m \times \mathbf{B}_{m}^{\mathrm{tot}}+\mathbf{d}_m \times \mathbf{F}_{m}^{\mathrm{mag}} ,
\end{equation}\label{torque}

where $\mathbf{B}_{m}$, $\mathbf{F}_{m}^{\mathrm{mag}}$ are the total magnetic field and the total magnetic force in the bead $m$, respectively. Because of their elongated shape, magnetic rods experience anisotropic viscous resistance. 
This effect is commonly represented through a friction tensor decomposed into components parallel and perpendicular to the rod axis,
\begin{equation}
\boldsymbol{\Gamma}_T
=
\zeta_{\parallel}\,\hat{\mathbf{s}}\hat{\mathbf{s}}
+
\zeta_{\perp}(\mathbf{I}-\hat{\mathbf{s}}\hat{\mathbf{s}}),
\end{equation}
where $\hat{\mathbf{s}}$ is the unit vector along the rod axis. For slender rods ($L/\sigma \gg 1$), hydrodynamic theory predicts $\zeta_{\parallel},\zeta_{\perp}\propto \eta_s L/\ln(L/\sigma)$, while corrections for moderate aspect ratios are commonly described using the semi-empirical expressions of Tirado and de~la~Torre \cite{Tirado1979,Tirado1984}. Including anisotropic friction is essential when the dynamical response of the system is of interest, particularly under time-dependent fields, as it directly affects relaxation times and synchronization behavior.
\begin{figure*}[t]
    \centering
    \includegraphics[width=0.7\textwidth]{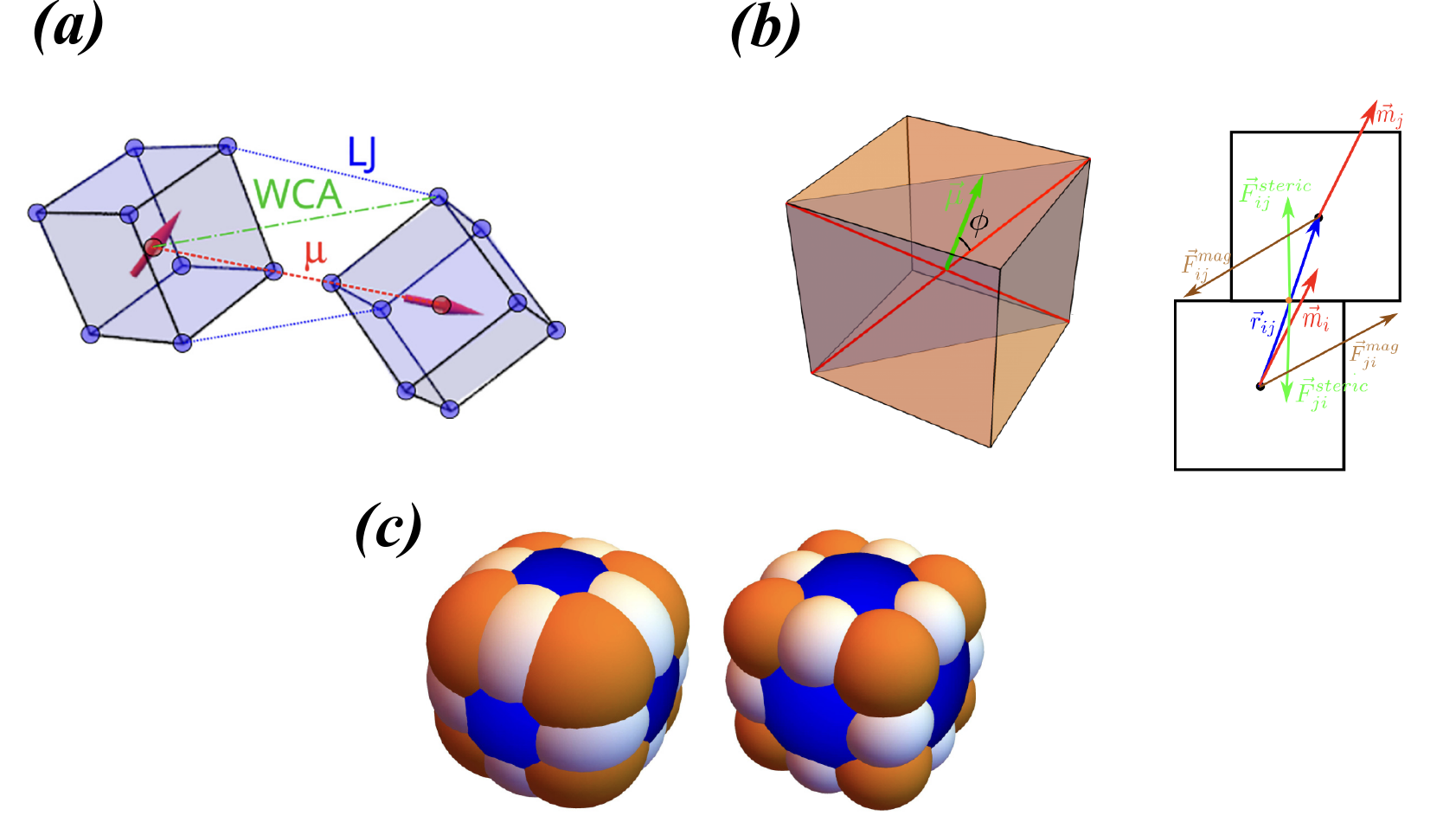}
    \caption{Modeling strategies for cubic magnetic colloids highlighting different levels of geometric representation. 
    (a) Discrete site-based model, where anisotropy arises from interaction sites located at corners, faces, or the particle center, enabling directional binding such as corner-, edge-, or face-contact configurations. Reproduced from Ref.~\cite{cub2} with permission from the American Physical Society. 
    (b) Illustration of contact-based steric modeling for polyhedral particles, where exact geometric overlap detection becomes increasingly complex and computationally demanding for realistic shapes and non-planar motion. Adapted from Ref.~\cite{PhysRevE.108.024601} with permission from the American Physical Society. 
    (c) Continuous shape representation using the superball model for increasing $q$ enabling systematic investigation of shape-dependent dipolar assembly. Adapted from Ref.~\cite{PhysRevE.105.024605} with permission from the American Physical Society.}
    \label{fig:cube_models}
\end{figure*}

Disk-like magnetic particles (platelets) have also been modeled using rigid multi-site representations in which the particle geometry is constructed from several steric interaction sites defined through a parametric representation of a circle in the particle reference frame, $\mathbf{r}_k = \mathbf{R} + R_p \left(\cos \phi_k\, \mathbf{e}_1 +\sin \phi_k\, \mathbf{e}_2 \right)$, and $\phi_k = \frac{2\pi k}{N}, \quad k = 1,\dots,N.$, while the magnetic interaction is described by a single central dipole moment \cite{platelts2023}. In this framework, forces acting on the interaction sites are transferred to a central rigid body, allowing the anisotropic excluded volume of the platelet to be captured without introducing distributed magnetic moments. Such models enable systematic investigations of dipolar assembly and collective dynamics in disk-like particles while keeping the magnetic description relatively simple. 

Cubic magnetic colloids constitute another class of anisotropic dipolar particles in which both particle geometry and internal dipole orientation break spherical symmetry. Experimental and numerical studies have shown that the orientation of the dipole relative to the particle shape strongly affects the resulting assembly pathways \cite{cub2,cub1,cub3,cub6}. In hematite nanocubes, for example, the permanent magnetic moment is tilted by approximately $12^{\circ}$ relative to the crystallographic axes, producing additional torques that favor kinked chains and non-collinear aggregates \cite{Brics2024}. Cubic particles can be modeled using discrete site-based approaches, where the particle geometry is constructed from a set of interaction sites located at corners, faces, or the particle center; see Fig. \ref{fig:cube_models}(a). In such models, anisotropy arises not only from the particle shape but also from the spatial distribution of interaction potentials. For instance, minimal coarse-grained descriptions employ Lennard-Jones interactions between corner sites, steric repulsion through center-corner interactions, and dipolar interactions located at the particle center, enabling the stabilization of corner-, edge-, or face-bound configurations depending on the balance of interactions \cite{cub2}. However, the accurate implementation of steric interactions for anisotropic polyhedral particles(Fig. \ref{fig:cube_models}(b)) remains a non-trivial task. Analytical contact-based models quickly become geometrically complex and computationally demanding, particularly for non-planar motion or realistic particle shapes \cite{PhysRevE.108.024601}.

To alleviate these geometric and computational difficulties, several studies employ continuous shape representations such as the \emph{superball} parameterization, see Fig.\ref{fig:cube_models}(c),
\begin{equation}
\left|\frac{2x}{a}\right|^{2q}+
\left|\frac{2y}{a}\right|^{2q}+
\left|\frac{2z}{a}\right|^{2q}\le1,
\end{equation}
which provides a smooth interpolation between spherical ($q=1$) and cubic ($q\to\infty$) shapes and enables systematic investigations of shape effects in dipolar assembly \cite{PhysRevE.105.024605,cub6}. This continuous representation allows one to explore how gradual changes in shape affect interaction anisotropy and collective behavior, providing a useful bridge between idealized models and experimentally realizable particle geometries.

Together, these modeling strategies illustrate that increasing geometric realism systematically enriches the range of collective behaviors that can be captured, but also increases computational cost and model complexity. The appropriate level of description must therefore be chosen based on the balance between physical accuracy and computational feasibility required for the problem at hand.
\section{Dipole-Particle Misalignment}
In many magnetic colloids the magnetic moment is not perfectly aligned with the particle symmetry. Such dipole-particle misalignment may arise either from a spatial displacement of the dipole relative to the particle center or from a tilt of the dipole with respect to the particle symmetry axis \cite{PhysRevE.110.064134}. From a modeling perspective, these two mechanisms introduce qualitatively different forms of anisotropy, as they modify both the symmetry of the interaction and the effective torque acting on the particles.

A general strategy to introduce such anisotropy is to decouple the magnetic dipole from the particle symmetry. In practice, this can be achieved either by displacing the dipole away from the particle center or by tilting it with respect to the particle symmetry axis, leading to non-central and non-collinear interactions. The particle dynamics is typically described within Brownian or Langevin dynamics frameworks, where forces and torques arising from dipolar interactions are integrated in time. Torques are typically evaluated using the anisotropic expression given in Eq.~\ref{torque}. In this context, dipole-particle misalignment introduces an additional coupling between translational and rotational degrees of freedom, which can significantly affect both steady-state configurations and dynamical response.

\begin{figure}[]
    \centering
    \includegraphics[width=0.8\linewidth]{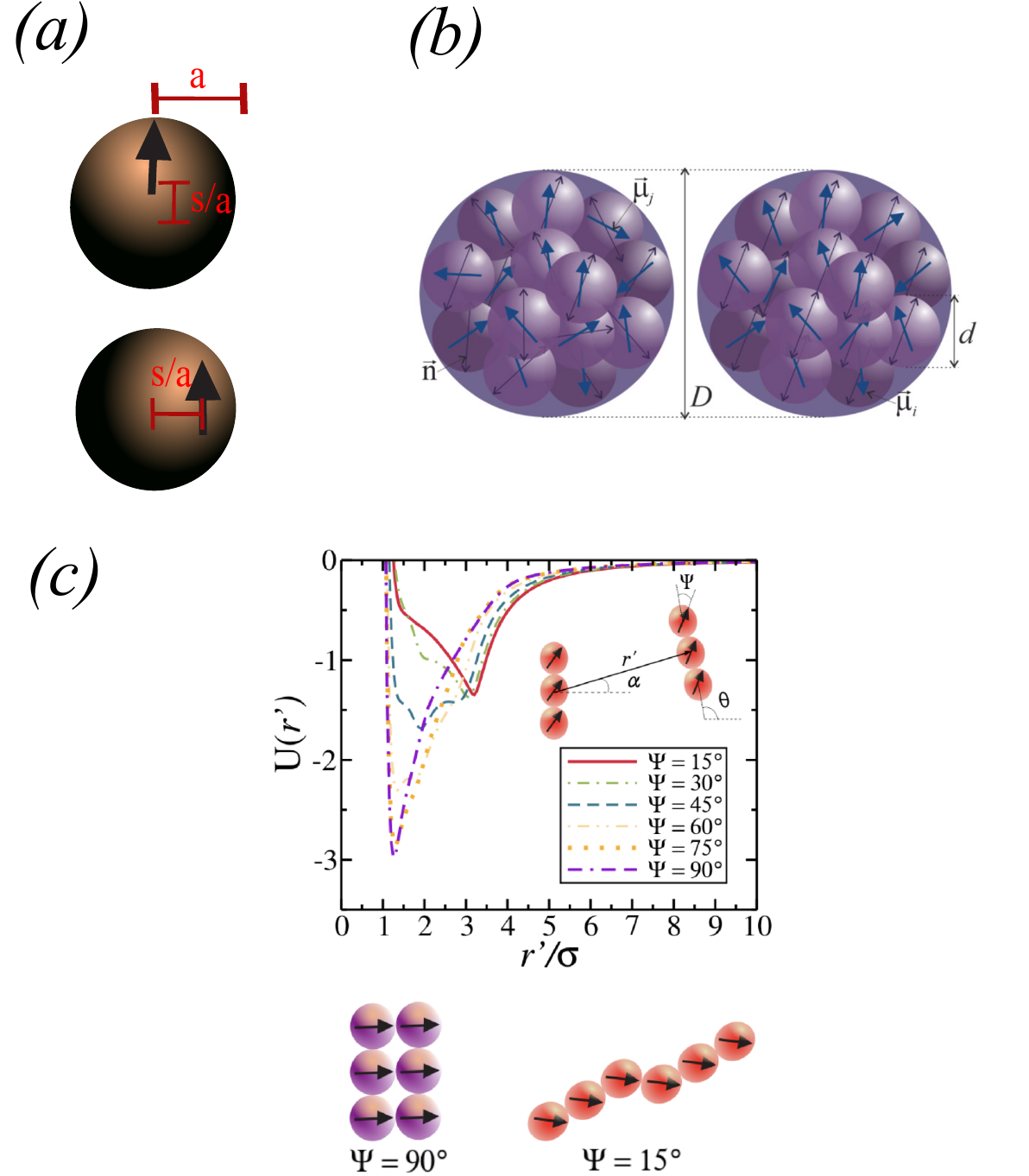}
    \caption{Modeling strategies for dipole-particle misalignment. 
    (a) Radial and lateral shifted-dipole models introducing off-centered magnetic moments. 
    (b) Multicore particle representation, where anisotropy emerges from multiple internal dipoles. Adapted from Ref.~\cite{PYANZINA2025126842}. 
    (c) Combined dipole misalignment and particle anisotropy in a multi-bead rod model, illustrating the dependence of the interaction profile on the misalignment angle $\Psi$. Adapted from Ref.~\cite{jorgeprsoft} with permission from the Royal Society of Chemistry.}
    \label{fig:misalignment_models}
\end{figure}

The simplest realization is the radially shifted-dipole (SD) model, where particles (often referred to as SD-particles) are represented as spherical cores carrying a point dipole located at $\mathbf{r}_m = \mathbf{r}_{\mathrm{cm}} + \mathbf{s}$, with $\mathbf{s}$ a fixed offset vector in the particle frame. The magnitude of this displacement is typically expressed in dimensionless form as $s = |\mathbf{s}|/a$, where $a$ is the particle radius (Fig. \ref{fig:misalignment_models}(a)). Interactions are computed via standard dipole-dipole potentials evaluated at the displaced positions, while steric repulsion is enforced via hard-sphere or soft-core potentials. This minimal model provides a controlled way to introduce dipole-particle misalignment, enabling systematic investigation of anisotropic interactions within a simple particle-based framework \citep{KANTOROVICH20111269,C1SM05186E}. 

However, the SD model remains a point-dipole approximation and does not account for the internal magnetization distribution of the particle. As a result, it cannot capture multipolar effects or spatially extended magnetic interactions, which may become relevant in particles with heterogeneous magnetic coatings or complex internal structure. In addition, Brownian dynamics simulations indicate that numerical stability imposes an upper bound on the radial displacement, typically $s \lesssim 0.8$, due to the proximity of the dipole to the particle surface\citep{radialshitvictoria,radialshiftmagneticfield,walker2026stockmayerfluidshifteddipole}. 

In contrast, lateral shift models introduce misalignment by displacing the dipole perpendicular to the particle symmetry axis (Fig. \ref{fig:misalignment_models}(a)), $\mathbf{r}_m=\mathbf{r}_{\mathrm{cm}}+\mathbf{s}_{\perp}$ with $\mathbf{s}_{\perp}\cdot\hat{\mathbf{u}}=0$ \cite{shearlaterally,D5ME00029G}. This modification breaks axial symmetry and leads to qualitatively different interaction landscapes, including the stabilization of compact aggregates, vesicle-like structures, and anisotropic clusters that are not accessible in centered-dipole systems. Compared to radial shifts, lateral displacements break axial symmetry more strongly and can lead to a more pronounced coupling between particle orientation and local structure, particularly in systems where directional interactions play a dominant role.

While most studies rely on explicit dynamical simulations, alternative approaches have been proposed to efficiently explore the large parameter space associated with dipole displacement. In particular, differential evolution-based methods have been used to predict the assembly of magnetic Janus particles with both radial and lateral shifts within a point-dipole framework, capturing equilibrium orientations and structural motifs under external fields while avoiding explicit time integration \cite{10.1063/5.0270900}. These approaches provide efficient access to equilibrium configurations, but do not resolve kinetic pathways or transient states, which remain essential for understanding field-driven assembly processes.

More refined models represent anisotropic magnetization by introducing multiple or effectively distributed off-centered dipoles attached to a steric core, mimicking magnetic caps or patchy coatings. In these models, interactions are computed as the sum of dipolar contributions between sites, allowing one to encode internal magnetic heterogeneity beyond the single-dipole approximation \cite{NOVAK2017214,zhu2025reconfigurable}. A particularly illustrative case is the double-dipole representation of Janus particles, where two oppositely oriented dipoles are embedded within a single particle. Such models generate effective multipolar interactions that promote lateral crosslinking and the formation of anisotropic network structures under external fields \cite{Schmidle2013}. 

A complementary strategy consists of explicitly resolving the internal magnetic structure using multicore models, where several permanent dipoles are embedded within a rigid particle, see Fig. \ref{fig:misalignment_models}(b). In this case, anisotropy emerges from the collective interactions between internal dipoles, providing a more realistic description of magnetic heterogeneity \cite{PYANZINA2025126842}. However, this increased level of detail significantly raises computational cost and introduces additional parameters related to dipole arrangement and coupling, limiting their applicability in large-scale simulations.

An alternative and often more practical approach consists of introducing misalignment through a tilt of the dipole relative to the particle symmetry axis. In this framework, the misalignment angle $\psi$ is defined through $\cos\psi = \hat{\boldsymbol{\mu}}\cdot\hat{\mathbf{s}}$, providing a direct parametrization of non-axial dipoles. This formulation is particularly convenient in simulations of anisotropic particles such as rods or ellipsoids, where the dipole orientation can be treated as an internal degree of freedom \cite{jorgeplos,jorgeprsoft} (Fig. \ref{fig:misalignment_models}(c)). Unlike shifted-dipole models, tilt-based approaches preserve the central position of the dipole while introducing angular anisotropy, allowing one to isolate the effect of directional misalignment without modifying the steric interaction geometry.

Anisotropic particles with non-axial dipole moments provide a natural realization of dipole-particle misalignment. Experimental \cite{PhysRevResearch.6.013287,PhysRevApplied.6.034002} and theoretical studies have reported systems such as ellipsoids, rods, and peanut-shaped particles exhibiting transverse or tilted magnetization under external fields \cite{peanut,peanut2}, as well as dipole-offset square particles displaying chiral interactions and asymmetric field responses \cite{dorsey2025chirality}. These systems highlight that dipole-particle misalignment is not merely a modeling abstraction, but a physically realizable mechanism that can be used to control self-assembly pathways and collective dynamics.
\begin{figure}[t]
\centering
\includegraphics[width=1.\columnwidth]{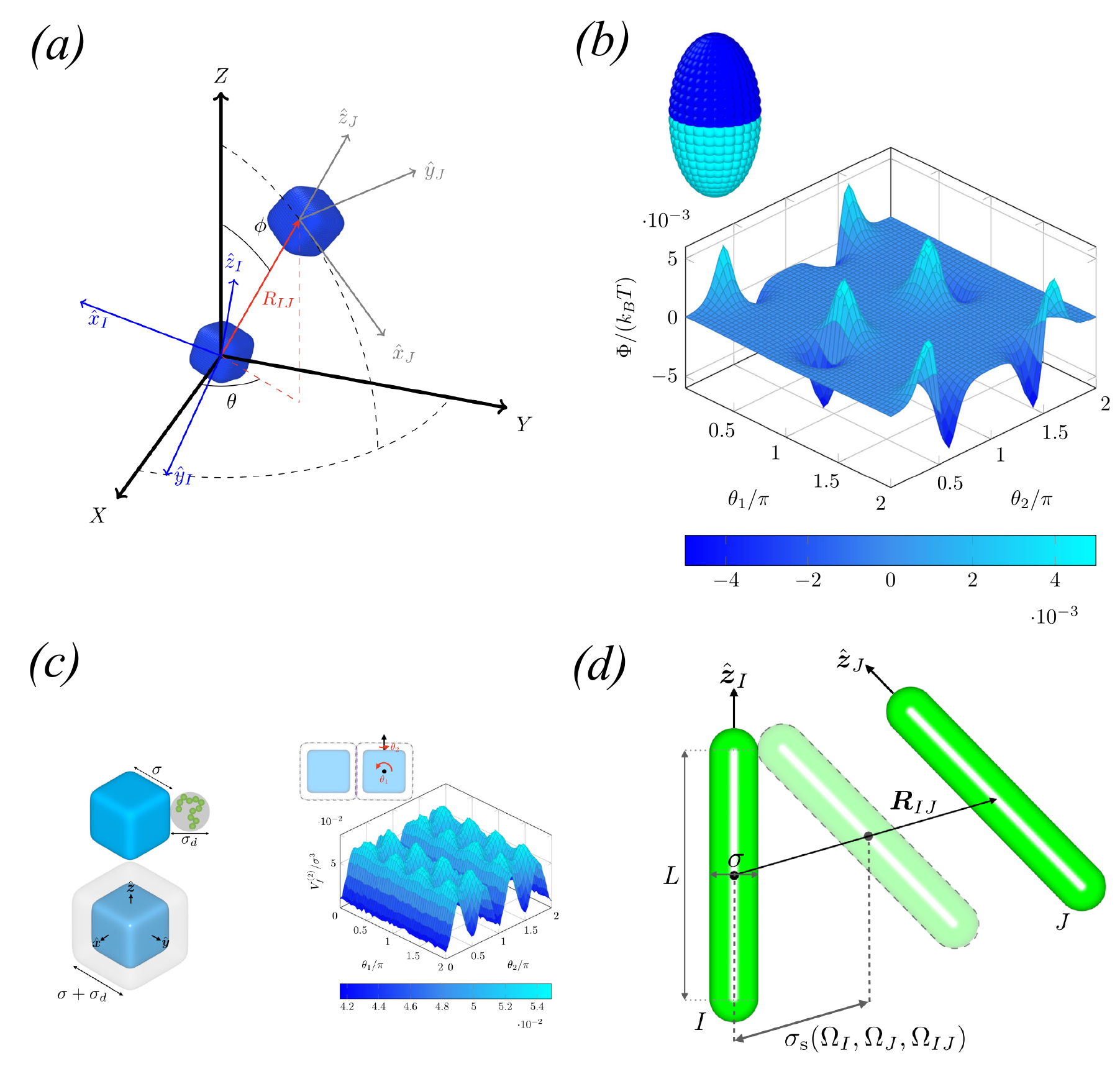}
\caption{Machine learning-based, physically informed descriptors for anisotropic coarse-graining.
(a) Parametrization of particle pairs in terms of relative position and orientational degrees of freedom.
(b) Orientation-dependent interaction energy landscape for anisotropic particles (ellipsoidal example).
(c) Example of faceted (cube-like) particles illustrating shape-dependent interactions.
(d) Example of elongated (spherocylinder-like) particles highlighting the role of geometry and orientation in the interaction model.
Adapted from Ref.~\cite{campos2024machine}, with permission from Springer Nature.
}
\label{fig:ml_descriptors}
\end{figure}

Overall, different modeling strategies for dipole-particle misalignment provide complementary levels of description. Shifted-dipole models offer a minimal and computationally efficient way to introduce anisotropy, while multi-dipole and multicore approaches capture internal magnetic structure at higher computational cost. Tilt-based formulations provide a simple alternative for anisotropic particles where the dipole orientation can be treated explicitly. The choice of approach therefore depends on whether the relevant physics is dominated by geometric asymmetry, internal magnetization structure, or directional coupling between particle orientation and dipolar interactions. 

\section{Emerging Frontiers: Machine Learning Approaches}
\begin{table*}[!t]
\caption{Overview of modeling strategies for anisotropic magnetic colloids.}
\makebox[\textwidth][l]{%
\hspace*{-0.8cm}%
\begin{tabular}{lccc}
\hline
Model & Magnetic description & Advantages & Limitations \\
\hline
Single-site (ellipsoids, Gay--Berne) & Single dipole & Computationally efficient & Limited shape realism \\
Multi-bead models & Distributed dipoles & Flexible geometry, realistic interactions & High computational cost \\
Shifted dipole models & Off-centered dipole & Simple anisotropy control & Limited internal structure \\
Multicore models & Multiple internal dipoles & Realistic magnetization & Very expensive \\
Machine-learned potentials & Learned effective interactions & High efficiency at large scale & Requires training data\\
\hline\label{tab:models}
\end{tabular}%
}
\end{table*}
Recent advances in machine learning (ML) have opened new avenues for the construction of coarse-grained interaction potentials, bypassing the often prohibitive computational cost associated with explicitly resolving interactions \cite{doi:10.1021/acs.jpcb.5c01451,https://doi.org/10.1002/adfm.202315177,doi:10.1021/acs.jcim.5c03035,TerRele2025MachineLM}. Within atomistic simulation frameworks such as molecular dynamics (MD), ML can be integrated either as a surrogate model for interactions or as a post-processing tool for structural and dynamical analysis. This integration improves predictive capabilities by increasing accuracy, decreasing computational demands, and revealing underlying patterns in complex, high-dimensional datasets \cite{stavrogiannis2026machine}. More recently, increasing attention has been given to anisotropic particles. In these approaches, the interaction potential is learned directly from fine-grained simulation or experimental data, $\Phi \approx \Phi_{\mathrm{ML}}$, using descriptors that encode both relative position and orientation. For instance, Argun \& Statt \cite{ml1} represented non-spherical particles as symmetry-preserving point clouds and benchmarked several ML models, showing that neuroevolution potentials (NEP) provide an optimal compromise between accuracy and efficiency, achieving order-of-magnitude speedups while accurately reproducing structural properties. A related earlier work \cite{10.1063/5.0206636} demonstrated that neural networks can directly map relative distance and orientation to energies, forces, and torques, bypassing explicit inter-site calculations and yielding speedups of up to one order of magnitude in Molecular Dynamics simulations of anisotropic particles. 

Complementary approaches focus on the construction of physically informed descriptors. Campos-Villalobos et al.~\cite{campos2024machine} introduced a bottom-up, data-driven framework in which anisotropic interactions are encoded through particle-centered structural descriptors that explicitly depend on both relative position and orientation. In this approach, the coarse-grained interaction is expressed as a linear combination of descriptors constructed from rotational invariants (so-called S-functions), allowing for a compact yet systematic representation of the full six-dimensional configurational space $(R_{ij}, \Omega_i, \Omega_j, \Omega_{ij})$ (Fig. \ref{fig:ml_descriptors}). 

By combining these descriptors with feature selection and simple regression schemes, the method enables the construction of single-site coarse-grained potentials that accurately reproduce fine-grained interaction landscapes, including strongly anisotropic and patchy systems. This strategy provides a connection between physically interpretable expansions of anisotropic interactions and data-driven coarse-graining approaches, while maintaining computational efficiency suitable for large-scale simulations. Similarly, extensions of SOAP descriptors (Smooth Overlap of Atomic Positions) to anisotropic particles (AniSOAP) provide a flexible representation of non-spherical particle interactions \cite{10.1063/5.0210910}, while alternative neural-network architectures incorporating geometric constraints have been shown to preserve key physical properties such as energy conservation and rotational invariance at significantly reduced computational cost \cite{PhysRevE.110.055305} .

Although these methods have been primarily developed for general anisotropic systems, they offer a promising pathway for magnetic colloids, where the combined effects of shape anisotropy, dipolar interactions, and external fields lead to highly complex and computationally demanding interaction landscapes. In particular, they could be used to learn effective interactions that incorporate both geometric anisotropy and magnetic contributions, such as dipolar coupling and field-induced alignment, within a unified framework. This is especially relevant for anisotropic magnetic particles, where the interplay between shape, dipole orientation, and external fields leads to complex interaction landscapes that are costly to resolve using conventional multi-bead or Ewald-based approaches. 

Beyond interaction modeling, ML has also been employed to identify emergent behavior in dipolar systems. For example, neural-network-based classification methods have been used to detect phase transitions in dipolar polymers, revealing previously unidentified structures without prior knowledge of the phase diagram \cite{perera2025confusion}. Despite these advances, challenges remain, particularly in constructing descriptors that consistently capture both geometric and magnetic degrees of freedom, and in extending current models beyond pairwise interactions. Nevertheless, ML-based coarse-graining provides a promising route toward bridging the gap between detailed particle-level models and large-scale simulations of anisotropic magnetic systems.

A summary of the main modeling strategies discussed in this review, together with their advantages and limitations, is provided in Table~\ref{tab:models}.
\section{Conclusions and Outlook}

The modeling of anisotropic magnetic colloids with permanent dipoles relies on a hierarchy of approaches, ranging from minimal single-site descriptions to multi-bead and multi-dipole representations. Rather than a single optimal framework, these approaches reflect different levels of resolution, each capturing specific aspects of the interplay between particle geometry, dipole orientation, and collective interactions. As a result, the choice of model is not universal, but must be guided by the physical mechanisms that dominate the system under consideration.

A central outcome of this review is that particle shape and dipole-particle misalignment act as primary control parameters of the interaction landscape. Even small deviations from axial symmetry can qualitatively alter effective interactions, leading to frustration, symmetry breaking, and distinct aggregation pathways. Consequently, models that neglect these features may fail to capture key aspects of the system behavior, particularly in anisotropic or field-driven regimes.

Despite significant progress, the field remains constrained by a fundamental trade-off between physical realism and computational efficiency. Long-range dipolar interactions require specialized numerical treatments, while increasing geometric and magnetic detail rapidly raises the computational cost. This limitation becomes particularly severe in dense systems or in simulations involving time-dependent external fields, where both accuracy and dynamical resolution are essential.

Machine learning approaches provide a promising route to alleviate this limitation by constructing effective interaction models that depend on both distance and orientation. However, their current applicability remains limited by the availability of reliable training data and by the difficulty of incorporating long-range and many-body effects in a consistent manner. In this sense, ML should not be viewed as a replacement for physically motivated models, but rather as a complementary tool within a broader modeling framework.

A key direction for future research is the development of hybrid approaches that combine coarse-grained physical models with data-driven corrections. Such strategies may enable simulations that retain the essential features of dipolar interactions while extending the accessible system sizes and timescales. In parallel, systematic validation against experimental systems will be necessary to establish the predictive power of these models and to identify the relevant observables that can be directly compared across different levels of description.

Overall, progress in this field will depend on the ability to clearly identify the appropriate level of modeling for a given problem, balancing geometric resolution, magnetic complexity, and computational cost. Advances along these directions are expected to play a central role in the predictive design and control of functional materials based on anisotropic magnetic colloids.

\section*{Declaration of competing interest}
The author declares no competing interests.

\section*{Acknowledgments}
The author acknowledges funding from the European Union’s Horizon
Europe ERA Fellowship project Pattern Formation in Magnetic Spinners (Pattspin, No. 101130777).

\end{document}